\documentclass[aps,pra,10pt,twocolumn,showpacs,superscriptaddress]{revtex4-1}
\usepackage[pdftex]{graphicx}
\usepackage{dcolumn}
\usepackage{bm}
\usepackage{svg}
\usepackage[utf8]{inputenc}
\usepackage[T1]{fontenc}
\usepackage{mathptmx}
\usepackage{physics}
\usepackage{chemformula}
\usepackage[normalem]{ulem}
\usepackage{textgreek}
\usepackage[caption=false]{subfig}
\usepackage{siunitx}
\usepackage{amsmath,amssymb}
\usepackage{xcolor}
\DeclareMathAlphabet{\mathcal}{OMS}{cmsy}{m}{n}
\usepackage[colorlinks=true, allcolors=blue]{hyperref}
\def\virgolette #1{``#1"}
\newsavebox{\measurebox}

\begin{document}

\title{High-speed Source-Device Independent Quantum Random Number Generator on a Chip}
\author{Tommaso~Bertapelle}
\affiliation{Dipartimento di Ingegneria dell'Informazione, Universit\`a degli Studi di Padova, via Gradenigo 6B, IT-35131 Padova, Italy}

\author{Marco~Avesani}
\affiliation{Dipartimento di Ingegneria dell'Informazione, Universit\`a degli Studi di Padova, via Gradenigo 6B, IT-35131 Padova, Italy}
\affiliation{Padua Quantum Technologies Research Center, Universit\`a degli Studi di Padova, via Gradenigo 6A, IT-35131 Padova, Italy}

\author{Alberto~Santamato}
\thanks{New affiliation: Nu Quantum Ltd., Broers Building, 21 JJ Thomson Avenue, Cambridge, CB3 0FA, United Kingdom}
\affiliation{Photonic Networks and Technologies Lab - CNIT, 56124 Pisa, Italy}

\author{Alberto~Montanaro}
\affiliation{Photonic Networks and Technologies Lab - CNIT, 56124 Pisa, Italy}
\affiliation{Scuola Superiore Sant'Anna, 56124 Pisa, Italy}

\author{Marco~Chiesa}
\affiliation{InPhoTec, Integrated Photonic Technologies Foundation, 56124, Pisa, Italy}
\affiliation{CamGraPhIC srl, 56124, Pisa, Italy}

\author{Davide~Rotta}
\affiliation{InPhoTec, Integrated Photonic Technologies Foundation, 56124, Pisa, Italy}
\affiliation{CamGraPhIC srl, 56124, Pisa, Italy}

\author{Massimo~Artiglia}
\affiliation{Photonic Networks and Technologies Lab - CNIT, 56124 Pisa, Italy}
\affiliation{Scuola Superiore Sant'Anna, 56124 Pisa, Italy}

\author{Vito~Sorianello}
\affiliation{Photonic Networks and Technologies Lab - CNIT, 56124 Pisa, Italy}

\author{Francesco~Testa}
\affiliation{Scuola Superiore Sant'Anna, 56124 Pisa, Italy}

\author{Gabriele~De~Angelis}
\affiliation{Photonic Networks and Technologies Lab - CNIT, 56124 Pisa, Italy}
\affiliation{Scuola Superiore Sant'Anna, 56124 Pisa, Italy}

\author{Giampiero~Contestabile}
\affiliation{Scuola Superiore Sant'Anna, 56124 Pisa, Italy}

\author{Giuseppe~Vallone}
\affiliation{Dipartimento di Ingegneria dell'Informazione, Universit\`a degli Studi di Padova, via Gradenigo 6B, IT-35131 Padova, Italy}
\affiliation{Padua Quantum Technologies Research Center, Universit\`a degli Studi di Padova, via Gradenigo 6A, IT-35131 Padova, Italy}

\author{Marco~Romagnoli}
\affiliation{Photonic Networks and Technologies Lab - CNIT, 56124 Pisa, Italy}

\author{Paolo~Villoresi}
\affiliation{Dipartimento di Ingegneria dell'Informazione, Universit\`a degli Studi di Padova, via Gradenigo 6B, IT-35131 Padova, Italy}
\affiliation{Padua Quantum Technologies Research Center, Universit\`a degli Studi di Padova, via Gradenigo 6A, IT-35131 Padova, Italy}

\begin{abstract}
A wide range of applications require, by hypothesis, to have access to a high-speed, private, and genuine random source.
Quantum Random Number Generators (QRNGs) are currently the sole technology capable of producing true randomness.
However, the bulkiness of current implementations significantly limits their adoption.
In this work, we present a high-performance source-device independent QRNG leveraging a custom made integrated photonic chip.
The proposed scheme exploits the properties of a heterodyne receiver to enhance security and integration to promote spatial footprint reduction while simplifying its implementation.
This characteristics could represents a significant advancement toward the development of generators better suited to meet the demands of portable and space applications.
The system can deliver secure random numbers at a rate greater than 20 Gbps with a reduced spatial and power footprint.
\end{abstract}
\maketitle

\section{Introduction}
Random Numbers Generators (RNGs) are one of the most influential tools of the current information age.
Many real-world applications like cryptography, simulations, gaming and fundamental physics experiments, already exploit such devices \cite{Herrero-Collantes2017}.
However, their capability to deliver genuine randomness is a matter of great concern, especially in cryptography: most of the known classical, quantum and post-quantum protocols demand, by hypothesis, to have access to an authentic random source.
Violating such a core assumption exposes the system to exploitable vulnerabilities which can completely break their security.
However, despite their key role for security applications, almost all the RNGs used nowadays are deterministic in nature: they rely either on computer algorithms or classical physics phenomena.
Hence, the stream of numbers can only mimic a random behavior and definitely cannot be considered a real source of randomness. 

Since randomness cannot arise from determinism, the probabilistic nature of quantum mechanics can provide a method to generate truly random numbers, thus rectifying one of the principal flaws of current cryptographic protocols \cite{Acin2016}.
Therefore, given the growing concerns about cyber-security, it is expected that the interest in such devices will increase in the near future \cite{qrng_report}.
Nonetheless, despite the premises, the un-avoidable non-idealities and imperfections concerning the implementation of such Quantum-based RNGs (QRNGs) must be carefully evaluated, especially when dealing with security applications \cite{Herrero-Collantes2017}.
Indeed, such non-idealities can be modeled as leaked information correlated with the QRNG itself (side information) and thus exploited to predict the generated numbers. 
From this point of view, QRNGs can be classified into three categories: Fully Trusted (FT), Device-Independent (DI), and Semi-Device Independent (Semi-DI). 

FTs are relatively simple to realize and provide some of the highest generation rates to date \cite{doi:10.1063/1.4922417, Bruynsteen2023}.
Nevertheless, everything concerning the system, including the imperfections of the hardware used, must be known and characterized, leading to generators with the lowest level of security.
A complete device characterization may be challenging, as it sets many working assumptions, none of which must be breached.
{For example, for CV protocols, which are relevant to our work, Smith et al. \cite{Smith2019} report an attack that could undermine the security of some of the said systems.}
On the counterpart, DI schemes offer the highest level of security since nothing concerning the device's physical implementation is trusted.
However, DI-QRNGs are extremely difficult to realize as they request a loophole-free Bell test and are several orders of magnitude slower than FT devices \cite{Pironio2010, PhysRevLett.111.130406, Bierhorst2018, PhysRevLett.120.010503}.
Therefore, although they are the best choice in terms of offered security, they are not feasible for real-world scenarios due to the limitations discussed above.
The Semi-DI approach, instead, aims to make a compromise between security and speed.
To achieve such, Semi-DI only trusts specific aspects of the device, for instance, the quantum source \cite{Cao_2015}, the measuring apparatus \cite{PhysRevA.90.052327, PhysRevA.97.040102, PhysRevLett.118.060503, PhysRevX.6.011020, Xu:16}, the underlying Hilbert space dimension \cite{Lunghi2014, Mironowicz2021}, the mean photon number \cite{Himbeeck2017semidevice,Rusca2019,Avesani2020d,Tebyanian2021,Tebyanian2021b} or the overlap bound \cite{Brask2017} of the emitted states.
Clearly, the Semi-DI strategy guarantees greater security since fewer assumptions are required to operate the QRNG, if compared with the FT framework, while still achieving several Gbps \cite{Avesani2018} in terms of generation rate, which is much higher than DI schemes and yet sufficient for current practical applications.

Alongside the particular protocol exploited to realize a QRNG, additional aspects must be also considered to target a wider range of use cases.
Aside from sheer performance, several constraints like cost, system integration, scalability, reliability, power and space consumption must be addressed.
In recent years, improvements in the integrated photonics field made such a technology an interesting solution for developing QRNGs capable of reaching the aforementioned goals.
Research on these engineered generators is active \cite{Abellan:16, Raffaelli_2018, Roger:19, Haylock2019, Tasker2021, Bai2021, Bruynsteen2023, Li2024, Wang_2024}, witnessing the transition from the bulkiness of the early prototypes to a newer generation able to better meet the demands set by real-world applications ranging from the Internet of Things (IoT) to satellite-based systems.

In particular, the requirements imposed by space applications are some of the hardest to achieve.
Yet, several emerging technologies, such as satellite Quantum Key Distribution (QKD) \cite{Liao2017}, demand on-board robust, secure and fast QRNGs, up to several Gbps \cite{Grunenfelder2020}.
Fulfilling these requirements in such a harsh, space and power-constrained environment is difficult with bulk devices.
Integrated photonics, on the other hand, has already demonstrated itself to provide several advantages, for example free-space and satellite QKD \cite{qcosone}, allowing to achieve the aforementioned goals.

This work proposes and realizes a high-performance source-device independent QRNG system designed to reduce encumbrance while promoting simplicity.
Such goals are achieved by leveraging a custom-made integrated photonic chip that operates at the standard $1550~\si{\nano\meter}$ telecom wavelength without any active feedback stabilization control loop, and that is mounted on a dedicated RF PCB with all the electronics required for its operation.
The generator exploits the protocol described in \cite{Avesani2018}, allowing the certification of private and secure randomness with no assumption concerning the quantum source.
Overall, the device combines high speed and security and reaches a generation rate greater than $20~\si{Gbps}$ with a reduced spatial footprint, which makes it the first source-device independent QRNG on-chip and currently the fastest Semi-DI QRNG.
In light of such, we believe our device represents a significant step forward for QRNG systems that can both aim for portable and space applications with high-speed QKD as the primary target.

The paper is structured as follows: Section \ref{sec:sdi_protocol} briefly outlines the source-device independent QRNG protocol used, Section \ref{sec:chip_design} provides details concerning the device design, fabrication and packaging, Section \ref{sec:experimental} describes the experimental setup adopted to characterize and estimate the QRNG performances, while Section \ref{sec:results} reports the obtained results.

\section{Heterodyne Source-DI protocol}\label{sec:2}\label{sec:sdi_protocol}
In the context of cryptographic-oriented QRNGs, the device must be resilient against an attacker, often known as Eve, whose aim is to predict the latter outcomes.
To do so, the malicious adversary will exploit every knowledge about the generator implementation and any classical or quantum resource available, known as side-information $E$, to maximize its probability of correctly guessing the outcome $X$ conditioned on $E$.

The source-device independent protocol exploited in this work divides the QRNG into two sub-systems: the quantum source and the {heterodyne receiver (measuring apparatus)}.
While the first is assumed to be totally controlled by the malicious adversary, the latter is trusted and fully characterized.
It is important to note that the receiver's trustfulness does not {preclude the attacker's awareness of its operations details}.
{Indeed, it is presumed that the adversary knows the set of POVMs (Positive Operator Values Measurements) implemented by the measuring apparatus.}
Therefore, to maximize the chances of correctly guessing the outcome $X$ at each round of the QRNG protocol, Eve prepares and sends to the receiver a state $\hat{\rho}_A$ that can be written as an incoherent superposition $\hat \rho_A=\int d\beta\, p\left( \beta \right )\hat{\tau}_\beta^A$ of states $\tau_\beta^A$ with probability $p\left( \beta \right)$.
As already shown in \cite{Avesani2018,Avesani2020c}, such a conditional guessing probability,
$P_\text{guess}\left( X\vert E \right)$, is upper-bounded by the following:
\begin{equation}
  P_\text{guess} \left( X\vert E \right) \leq \max_{\hat{\Pi}_A,\, \hat{\tau}_A} \Tr[\hat{\Pi}_A \hat{\tau}_A]\, ,
  \label{eq:EvePguess}
\end{equation}
with $\hat{\Pi}_A$ the discretized version of the heterodyne's POVM.
Thus, given the receiver’s resolution $\delta_q$, $\delta_p$, and that $\Tr[\hat{\Pi}_A\, \hat{\tau}_A]\leq 1\text{/}\pi$ for every quantum state $\hat{\tau}_A$ injected, the previous bound becomes:
\begin{equation}
  P_\text{guess} \left( X\vert E \right) \leq \max_{\hat{\Pi}_A,\, \hat{\tau}_A} \Tr[\hat{\Pi}_A \hat{\tau}_A] \leq \dfrac{\delta_q\delta_p}{\pi}\, .
  \label{eq:PguessUpperBound}
\end{equation}
{Such a} $P_\text{guess}(X\vert E)$ {bound} forbids the attacker to correctly guess the heterodyne outcome (raw randomness) with a probability greater than $\delta_q\delta_p\text{/}\pi$ even with complete knowledge of the system and holding the side-information $E$.
{Consequently, the amount of truly random bits that can be distilled from each measurement, quantified by the quantum conditional min-entropy $H_\text{min}(X \vert E)$, are:
\begin{equation}
    H_\text{min}(X \vert E) = -\log_2 P_\text{guess}(X \vert E) \geq -\log_2 \dfrac{\delta_q \delta_p}{\pi}\, .
\end{equation}
}

The analysis just outlined is valid for individual attacks.
However, it can be extended to encompass the more general coherent scenario, as shown in \cite{Avesani2018}.
Considering that a collection of QRNG measurement outcomes can be written as $\hat{\Pi}_{x_1}\otimes \hat{\Pi}_{x_2}\otimes \dots \hat{\Pi}_{x_n}$ and that the attacker can now prepare a general n-partite state $\hat{\rho}^{(n)}$, the latter guessing probability becomes:
\begin{equation}
    \begin{split}
        P_\text{guess}^{(n)} \left( X \vert E \right) &= \max_{x_i}\,  \max_{\hat{\rho}^{(n)}} \Tr[\hat{\Pi}_{x_1}\otimes \hat{\Pi}_{x_2}\otimes \dots \hat{\Pi}_{x_n} \hat{\rho}^{(n)}] = \\
        &= \prod_{n=1}^n\, \max_{x_i, \hat{\rho}_i}\Tr[\hat{\Pi}_{x_i}\hat{\rho}_i] \, .
    \end{split}
\end{equation}
This is because maximization over $\hat{\rho}^{(n)}$ corresponds to the maximum eigenvalue of the tensor-product operator, which is composed of semi-definite positive Hermitian operators.
Notice that in the last equality, the term inside the product is the single-shot $P_\text{guess}\left( X \vert E \right)$ valid for individual attacks.
Hence, the overall quantum conditional min-entropy is:
\begin{equation}
    H_\text{min} ^{(n)}\left( X \vert E \right)=n\, H_\text{min} \left( X \vert E \right)\, .
\end{equation} 

{Finally, to} extract such randomness, a universal hashing function compliant with the constraints dictated by the "\textit{Leftover Hashing Lemma}" \cite{RennerLeftOver} can be adopted.
\begin{figure*}[htbp]
    \centering
    \vspace{5mm}
    \includegraphics[width=1.3\columnwidth]{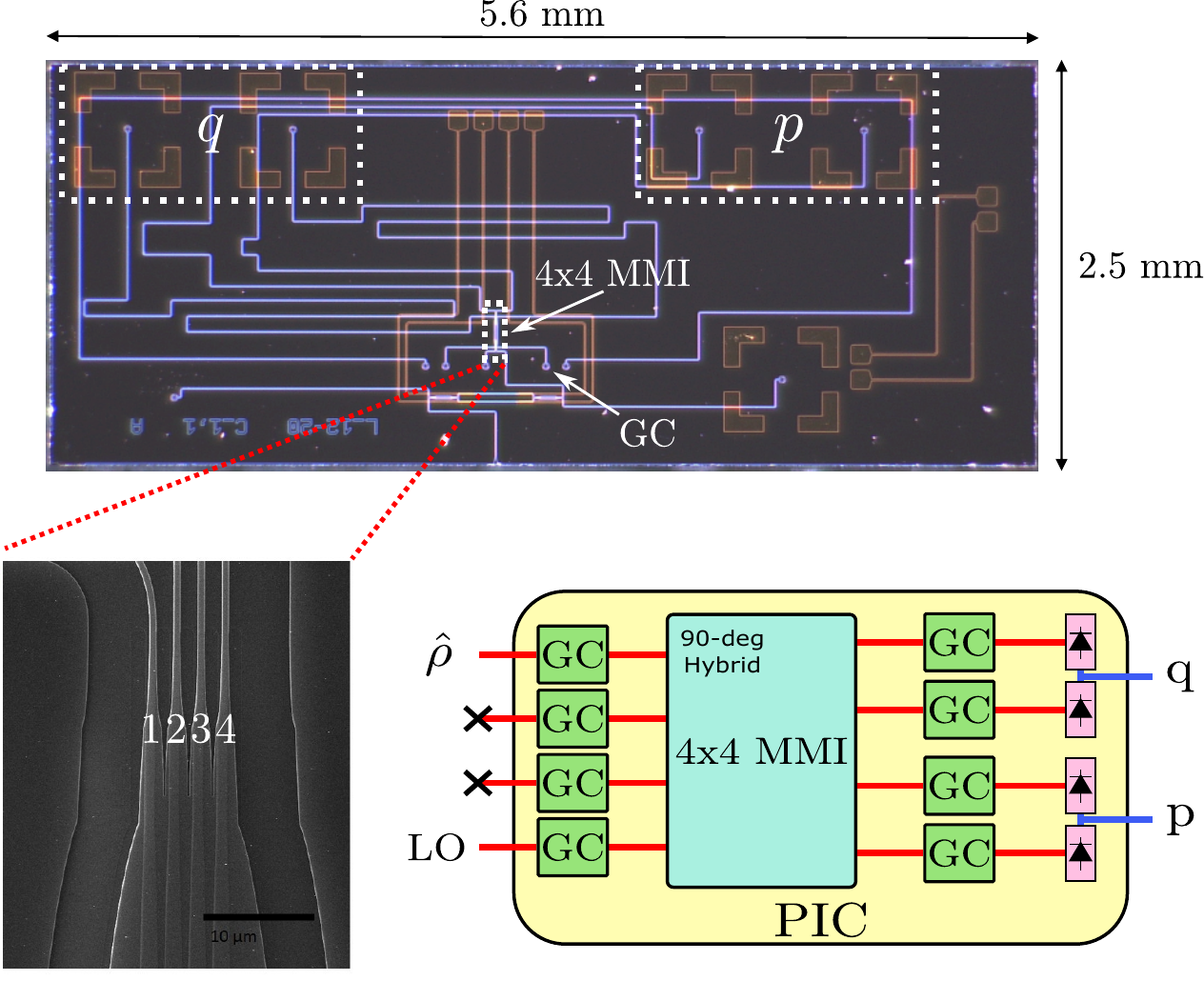}
    \caption{
    Image of the integrated photonic circuit implementing the heterodyne receiver.
    The 4x4 MMI can be accessed through grating couplers with the four outputs, $p$ and $q$, indicated with dashed lines.
    The routing between the latter grating coupler and the 4x4 MMI is done with waveguides of the same length and number of curves.
    {The bottom left inset shows a zoom on the 4x4 MMI outputs; the shallow etch can be noticed.
    The bottom right inset displays a high-level depiction of the PIC functioning as a heterodyne receiver.
    Notice that the photodiodes are hybridly integrated in the PIC.}
    }
    \label{fig:chip}
\end{figure*}

\section{Design, Fabrication and packaging of the integrated chip}\label{sec:3}\label{sec:chip_design}
\begin{figure*}[htbp]
    \centering
    \includegraphics[width=1.4\columnwidth]{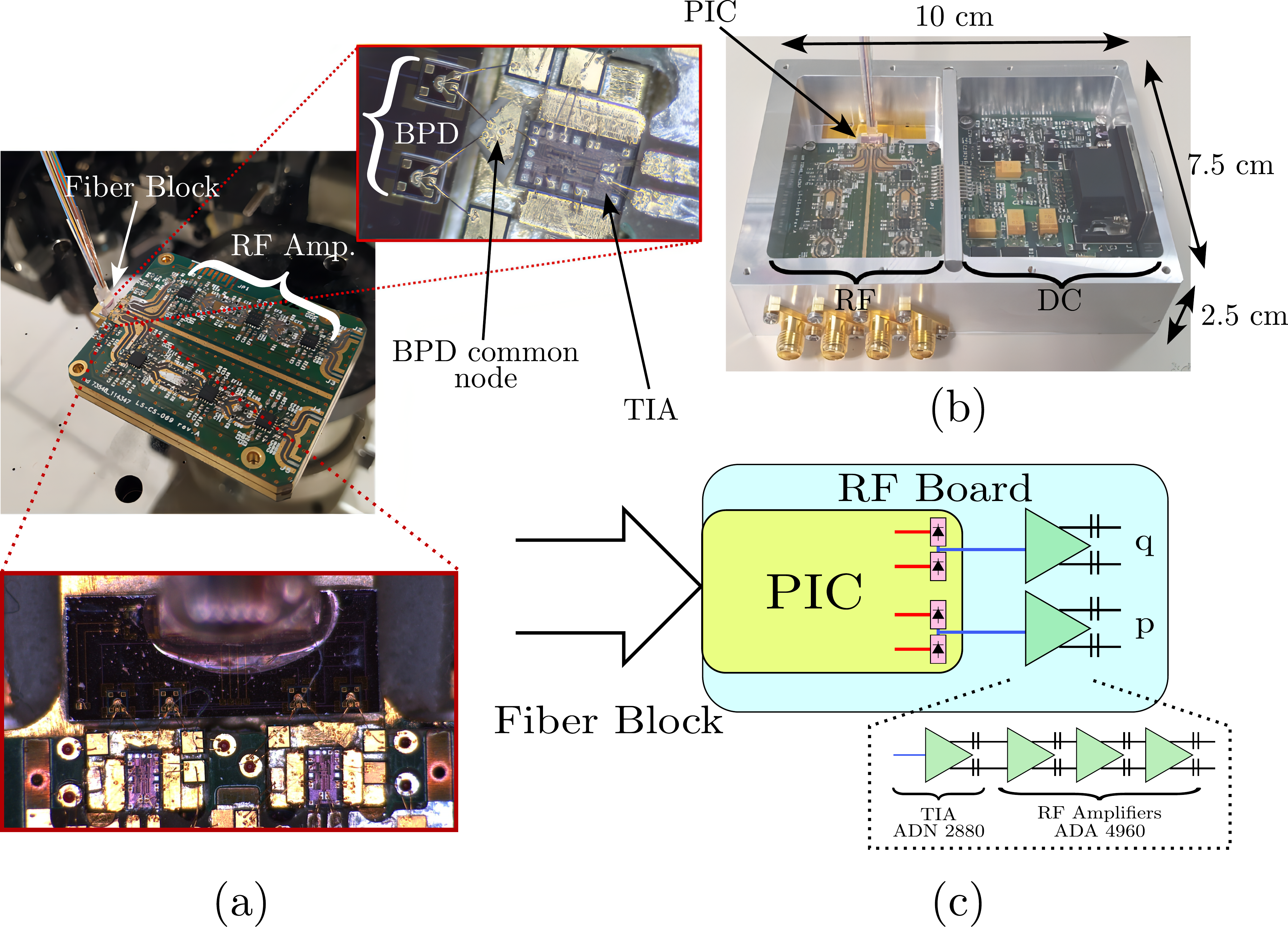}
    \caption{a) Assembly of the PIC with the RF electronics and fiber block for optical input.
    {The upper inset displays a close-up view of two hybrid-integrated photodiodes forming a balanced photodiode.
    The latter is connected to the transimpedance amplifier through short wire bonds.
    In the bottom inset, a zoom of the fiber block pigtailed on the PIC.}
    b) Complete assembly, comprising also the DC control, inside the metallic package.
    {High-level block diagram of the interconnection between the fiber block, PIC, and RF board.}
    }
    \label{fig:assembly}
\end{figure*}

An image illustrating the Integrated Photonic Chip (PIC) implementing the heterodyne receiver is presented in Figure {(\ref{fig:chip})}.
The PIC is fabricated using InPhoTec $220~\si{\nano\meter}$ SOI technology, providing $220\text{x} 480~\si{\nano\meter}$ photonic waveguides for optical routing.
{The heterodyne quadratures $q$ and $p$ are separated using a 90-deg optical hybrid realized with a 4x4 Multi-Mode Interferometer (MMI), which mixes the quantum signal with an intense coherent state.
It is important to note that since silicon photonics cannot integrate these sources directly into the chip substrate,
Consequently, hybrid integration or external light provision is necessary.
Although the former offers a smaller footprint, it presents numerous challenges, such as the requirement for exact sub-micrometer alignment between the source and the PIC.
Nevertheless, to demonstrate our chip's optimal performance, we chose to avoid the complexities associated with hybrid integration.}
{The} quadratures are than detected by a pair of balanced photodiodes.
Compared to 90-deg optical hybrids realized with optical splitters and phase shifters, a 4x4 MMI allows us to obtain a smaller circuit that does not need any active phase control \cite{Halir:11}.
However, the absence of the latter must be compensated with a robust technology that can guarantee the fabrication of MMIs with minimum phase error between the four channels.
Additionally, this technology must also maintain minimal differential losses among these channels.
For this reason, we designed a shallow-etched ridge waveguide-based MMI, see Fig. {(\ref{fig:chip}) bottom left inset}, to ensure low deviations from specifications \cite{Halir:11}.
Sixteen dedicated interferometric test structures were used to characterize the fabricated MMIs.
The results revealed a mean phase spacing between the four channels of $\sim 89.99~\si{\degree}$ with a standard deviation of $3.24~\si{\degree}$.
One test structure also underwent thermal variations analysis, with temperatures spanning from $15~\si{\degree}\text{C}$ to $80~\si{\degree}\text{C}$.
Findings indicated a maximum deviation of $<6\%$ at $80^\circ \text{C}$ compared to standard $25^\circ\text{C}$ room temperature, while $<3\%$ degrees deviations for temperatures below $60^\circ\text{C}$.
The optical inputs and outputs of the PIC were realized with Grating Couplers (GCs). 
Particular attention was given when routing each MMI's output with its corresponding GC.
As can be seen from Fig. {(\ref{fig:chip})}, the $p$-path 1-4 and $q$-path 2-3 have identical length and curve counts to minimize delays and relative losses.
Moreover, optical crossings were avoided to prevent cross-talking effects.
GCs and waveguides losses were estimated by fabricating and evaluating dedicated test chips.
Measurements revealed an average coupling loss between GCs and single-mode fibers of $4~\si{dB}$, and an average waveguide propagation loss of $\sim 2~\si{dB}\text{/}\si{\centi\meter}$.
The latter was estimated with the cut-back method \cite{Keck:72}.
The four MMI inputs are accessed with a single-mode fiber array pigtailed on the chip and coupled to the input GCs, see Fig. (\ref{fig:assembly}a).
{To align the former fiber block to the MMIs input GCs, the latter are arranged linearly between two extra GCs connected back-to-back, forming a loop. This loop is commonly used for aligning fiber blocks onto photonic integrated circuits \cite{Zimmermann_2011, Marchetti_2019}.
Specifically, during the process, light is injected through the fiber array to one of the two GCs composing the loop while connecting the other to a power meter.
Then, a six-axes, three linear and three rotational, industrial fiber alignment system exploits such feedback to optimize the fiber's block position.
Once aligned, a polymeric glue (also acting as a refractive index matching material) is dispensed and cured with UV (Ultra Violet) light to hold the fiber array in position.}
{Of the aforementioned MMI inputs, one of such serves as the} heterodyne Local Oscillator (LO) optical inlet.
{To convert the extracted quadratures in electrical signals, two} couples of high-speed photodiodes (PDs) (\textit{Kiosemi KPDEH20LC}) with responsivity $\sim 0.8~\si{\ampere}\text{/}\si{\watt}$ were hybridly integrated with the chip using a pick-and-place technique at the level of output GCs (Fig. (\ref{fig:assembly}a) top zoom) via vertical coupling.
Each PD pair was electrically connected to form a balanced photodiode (BPD) as detailed in the zoom on the top of Fig. (\ref{fig:assembly}a).
Hybrid integration was necessary due to the unavailability of active devices within our in-house technology.
Based on simulations, a nominal coupling loss of $2.5~\si{dB}$ between the PDs and GCs was expected.
By also incorporating the alignment tolerances due to PDs to GCs misalignment ($\pm 5 ~\si{\micro\meter}$) during integration, the coupling loss was estimated not to exceed $3~\si{dB}$.
This loss can be almost eliminated using integrated photodiodes.
Simulations also indicated a maximum optical power unbalance of $0.5~\si{dB}$ between the two PDs composing each BPD.

The chip, alongside its fiber array and PDs, was then assembled on a custom-made RF PCB, which houses the electronics for readout, as shown in Fig. (\ref{fig:assembly}).
The interface design between the PIC and the external electronic circuit is crucial to ensure proper signal amplification and conditioning when working with bandwidths in the $\si{\giga\hertz}$ range.
In the present system, this interface comprises the balanced photodiodes (BPDs) on the PIC end and low-noise transimpedance amplifiers (TIAs) on the RF PCB side.
The TIAs serve as the initial amplification stage, boosting the BPD currents.
TIAs and BPDs were interconnected through ball-wedge wire bonds of $\sim 700~\si{\micro\meter}$.
Balanced detection of the two quadratures is achieved by wiring the cathode of one PD of the couple with the anode of the other, as illustrated in the top zoom of Fig. (\ref{fig:assembly}a).
Since high-speed operations necessitate broad bandwidths, a higher degree of noise is introduced compared to lower-frequency systems. Such a noise increase degrades the receiver's sensitivity.
Moreover, wire bonds introduce parasitic inductances that may limit the operating bandwidth while generating unwanted oscillations of the active electronic components \cite{7953542}.
Therefore, considering all the aforementioned constraints, the interface was designed to best compromise between noise, bandwidth, and gain values to approach shot-noise limited measurements.
\begin{figure*}[t]
    \centering
    \includegraphics[width=0.76\linewidth]{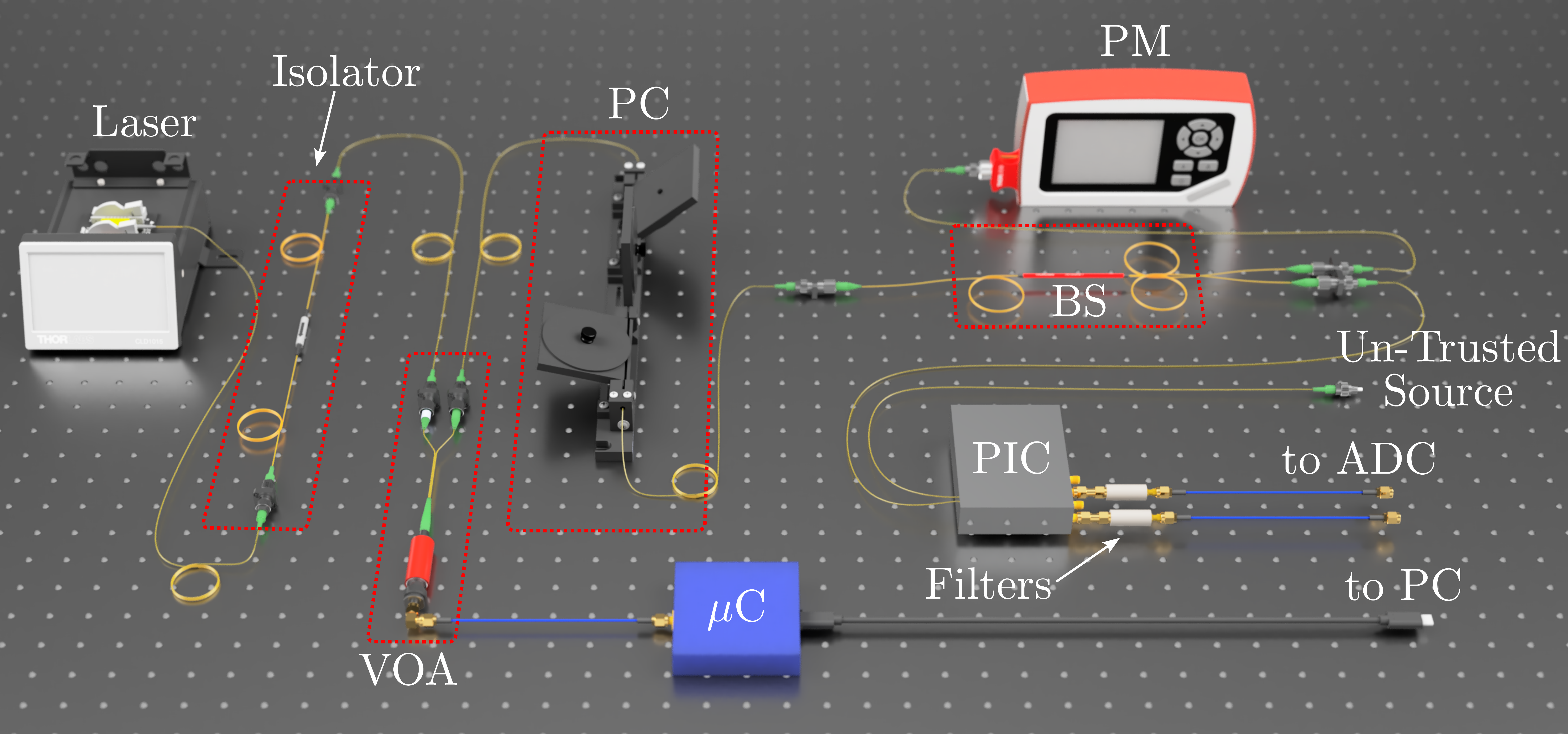}
    \caption{Schematic representation of the experimental setup used. The latter consists of a $1550~\si{\nano\meter}$ laser acting as a Local Oscillator (LO) for the optical heterodyne receiver.
    The isolator prevents back-reflections causing laser instabilities, the Polarization Controller (PC) aligns the LO polarization before coupling it into the Photonic Integrated Circuit (PIC), the Variable Optical Attenuator (VOA) is used to scan the LO power during the calibration phase to estimate the conditional min-entropy parameter, and through the Beam Splitter (BS) (99:1 splitting ratio) the optical power injected into the PIC is determined.
    {The device's RF signals are provided as differential pairs. For each of such, one is terminated to $50~\si{\ohm}$, and the other is filtered with analog filters and digitized by an Analog-to-Digital Converter (ADC) connected to the PC. This is because the ADC used only accepts single-ended signals. Moreover, the PC controls everything concerning the calibration to automatize the procedure.}}
    \label{fig:setup}
\end{figure*}
The TIAs (\textit{Analog Devices ADN 2880}), depicted in the inset of Fig. (\ref{fig:assembly}a), feature a bandwidth of $2.5~\si{\giga\hertz}$, $4400~\si{\ohm}$ differential transimpedance gain, and an input referred noise of $315~\si{\nano\ampere}$.
Each TIA was then coupled to a chain of amplifiers to match the subsequent analog-to-digital converter voltage range.
The latter amplification chain frequency response was oversized to preserve the operating frequency window set by the TIAs and minimize classical correlations. Specifically, the amplification chain is three-staged, each of which (\textit{Analog Devices ADA 4960}) has a bandwidth of $5~\si{GHz}$.
This choice made negligible impact on the receiver's SNR since this last is dominated by the TIAs noise as by Friis's formula \cite{1695024}.
The amplification system, including the TIA, provides an overall differential gain of $281 600~\si{\ohm}$.
Calculations indicated a total input referred electronic noise level of $\sim 6 ~\si{\pico\ampere}\text{/}\sqrt{\si{\hertz}}$ when neglecting the small contribution of the photodiode's dark current and Johnson noise.
From this, it was determined that the {clearance} is upper limited by $11~\si{dB}$ when $1~\si{\milli\watt}$ {reaches} each photodiode{, which translates to roughly $20~\si{\milli\watt}$ at the heteordyne's input after considering insertion losses and the power splitting contribution of the MMI}.
Such an estimation presumes that the MMI outputs are perfectly balanced, the PDs' responsivity deviates minimally from the nominal value of $0.8 ~\si{\ampere}\text{/}\si{\watt}$, and the PDs/GCs are perfectly aligned.
The $11~\si{dB}$ upper-bound is consistent with the experimental value of $\sim 9~\si{dB}$ reported in Section \ref{sec:4}.
Discrepancies are due to violations of the ideal working hypothesis discussed above.
A DC electronic board was designed and manufactured to manage and supply the necessary electrical biases for the RF electronics and BPDs.
Finally, as shown in Fig. (\ref{fig:assembly}b), the PIC, the RF PCB, and the DC control electronics were integrated inside a metallic enclosure to ensure EMI protection, insulate the heterodyne receiver from external disturbances, and maintain the high sensitivity level needed to detect quantum signals.
The fully assembled module, shown in Fig. (\ref{fig:assembly}c), outputs two pairs of differential electrical signals ($q$ and $p$) via SMA connectors.
The module's overall dimensions are $7.5\text{x}10\text{x}2.5~\si{cm}$, as indicated in Fig. {(\ref{fig:assembly}b)}.

\section{Experimental implementation}\label{sec:4}\label{sec:experimental}
A schematic representation of the experimental setup used to run the heterodyne-based source-device independent QRNG protocol is presented in Fig. (\ref{fig:setup}).
The LO is an external $1550~\si{\nano\meter}$ Distributed FeedBack (DFB) laser, whose power and polarization are controlled by a Polarization Controller (PC) and a Variable Optical Attenuator (VOA) before the injection of the former light into the photonic chip.
Subsequently, a Beam Splitter (BS) routes 1\% of the LO power to an optical Power-Meter (PM, Thorlabs PM100D with S122C) connected to a computer, while the remaining 99\% is sent to the heterodyne receiver.
The power-meter and the VOA are required only during the calibration phase in which $H_\text{min} \left( X\vert E \right)$ is estimated.
The PC, instead, is necessary to maximize the amount of optical power coupled into the PIC due to its sensitivity to polarization.
{The integrated device then performs a heterodyne measurement on the incoming quantum state, yielding two pairs of RF differential signals with a bandwidth of $2.5~\si{\giga\hertz}$ each.
To interface it with a single-ended RF chain and ADC, we kept one of the channels while terminating the other to $50~\si{\ohm}$.
}
Finally, these are filtered by an analog high-pass filter with a $48~\si{\mega\hertz}$ cutoff frequency to eliminate low-frequency spectral noise from the LO, sampled by an oscilloscope at a rate of $25~\si{GSps}$ with an $8\text{-bit}$ resolution per channel and transferred to a computer {off-line} for digital processing.
The latter step accounts for a pre-processing procedure, based on \cite{Avesani2018}, that extracts the portion of the spectrum of the measurements for random number generation and random extraction.

Before operating the QRNG, {the heterodyne receiver must be characterized and the pre-processing parameters defined}.

\begin{figure}[t]
    \centering
    \includegraphics[width=\linewidth]{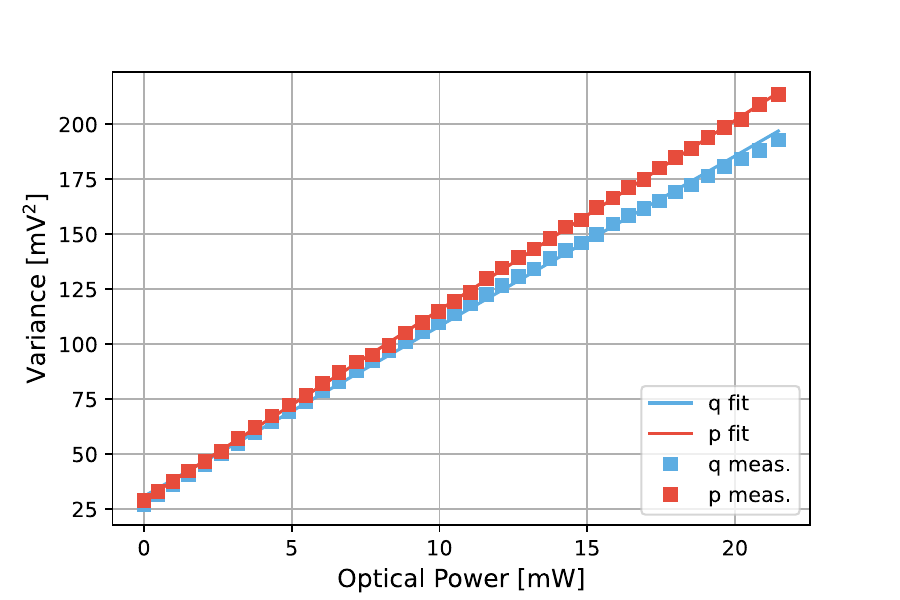}
    \caption{Relation between the LO optical power and the measured signal variance for both processed RF channels, representing the $q,p$ quadratures.
    The plot shows a linear relationship between the two quantities in all the LO power range, with a similar trend for both channels.
    The parameters of the linear fit are used in the calibration procedure to convert from digitized units to vacuum units (VU).}
    \label{fig:Calibration}
\end{figure}

{The former is achieved when its phase-space resolution $\delta_q$ and $\delta_p$ is estimated.}
To do this, we observe that the photo-current associated with each quadrature is proportional to the local oscillator's power $P_{LO}$.
Therefore, if the system behaves linearly with respect to such a power, then the ADC measures can be re-scaled from their original units $V_{q,p}$ into the phase-space vacuum units with a constant value $k_{q,p}$.
To determine the latter, we injected the vacuum state into the device's signal port while scanning $P_{LO}$ from zero up to the maximum applied value of $21.15 \pm 0.01~\si{\milli\watt}$.
For every local oscillator's power used, we acquired $10$ million data-points for both quadratures, and computed their variance $\sigma_{V_{q,p}}^2$.
The resulting $P_{LO}$-$\sigma_{V_{q,p}}^2$ trend, see Fig. (\ref{fig:Calibration}), exhibits a linear behavior.
For each quadrature, the re-scaling factor is given by $k_{q,p}=2\, m_{q,p} P_{LO}$, with $m_{q,p}$ the angular coefficients of the $P_{LO}$-$\sigma_{V_{q,p}}^2$ fitted straight lines.
The $\delta_{q,p}$ values can be then obtained by re-scaling the measurements' resolution, which is set by the ADC range and bit-width, with the retrieved $k_{q,p}$, as described in \cite{Avesani2018}.
From the data reported in Fig. (\ref{fig:Calibration}), the estimated $m_{q,p}$ coefficients are different.
The discrepancy is mostly due to {LO power unbalance between the two homodynes measuring the $q$ and $p$ quadrature,} variations and tolerances in the detector's responsivity and the system's amplification gain.
Nevertheless, the calibration purpose is to consider such variations to map the measurements in vacuum units {, and compensate for such imperfections}.
Hence, we highlight that the different $m_{q,p}$ do not represent a problem from the protocol's perspective; they only capture the responses of the photodetection systems for proper rescaling.
{Yet, some imperfections remain challenging to address, and a refined heterodyne receiver model is essential for assessing the impact on performance, such as the non-perfect orthogonality between quadrature measured.
This is why we required the implementation of the MMI as a 90-deg hybrid to minimize its deviation from ideality, as reported in Section (\ref{sec:3}).
Such a requirement should guarantee negligible effects on performance, but its theoretical inclusion in a security proof is an area of ongoing research, and we are planning to incorporate it into our methods in future works.}
{It is} worth noticing that with the calibration procedure adopted, the electronic noise, assumed classical and thus predictable, is un-trusted.
The approach makes the random estimation more conservative but secure against an attacker's manipulation of such a noise.

\begin{figure}[t]
    \centering
    \includegraphics[width=\linewidth]{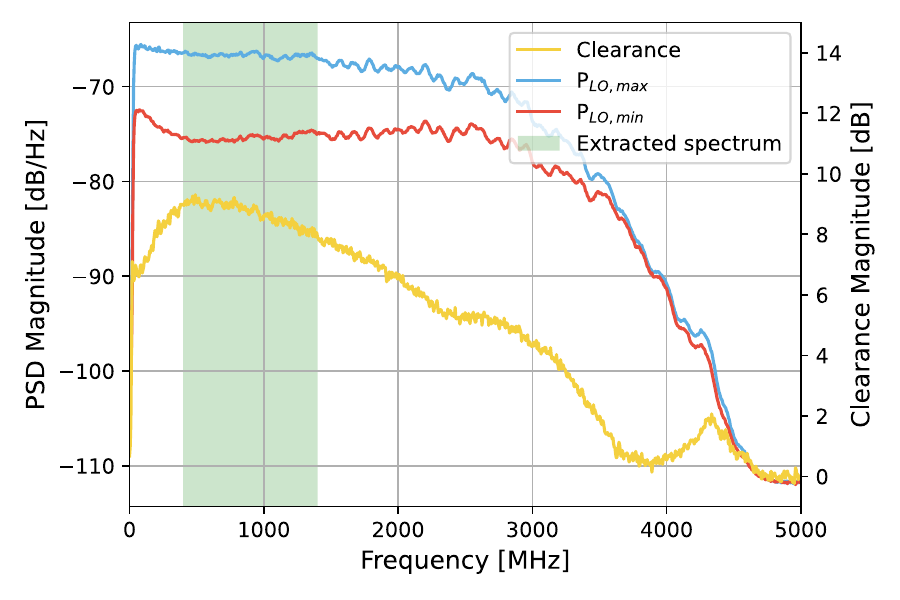}
    \caption{{Typical heterodyne receiver's clearance, in yellow, and PSDs (before digital pre-processing) either with no and maximum (21 mW) LO power respectively in red and blue.
    Notice that only of the two channels is reported for reference, since performances are similar for both.} Notice that the device clearance is greater than $4~\si{dB}$ within the entire operational bandwidth ($2.5~\si{\giga\hertz}$) reaching $\sim\, 9~\si{dB}$ at its maximum.
    Moreover, the figure also highlights (in green) the portion of the spectrum exploited for the random extracting procedure during the experiments.}
    \label{fig:PSD_Clearance}
\end{figure}

{Following the characterization of the receiver, the local oscillator power $P_\text{LO}$ was established, and consequently the pre-processing parameters were defined.
Since the generation rate increases as $\delta_{q,p}$ decreases, $P_\text{LO}$ was set to its maximum value ($21~\si{\milli\watt}$).
Then, from the heterodyne measurements, their Power Spectral Densities (PSDs) and clearance were computed; see Fig. (\ref{fig:PSD_Clearance}) for reference.
As can be seen, we decided on a spectral window ranging from $400$ to $1400~\si{\mega\hertz}$.
Such a choice rejects the technical noise present at lower frequencies, is sufficiently far from the system cutoff frequency ($2.5~\si{\giga\hertz}$ at $-3~\si{dB}$) and ensures that the signals used for the random extraction possess a clearance of at least $8~\si{\dB}$.}

\section{Results}\label{sec:5}\label{sec:results}
\begin{figure}[t]
    \centering
    \includegraphics[width=\linewidth]{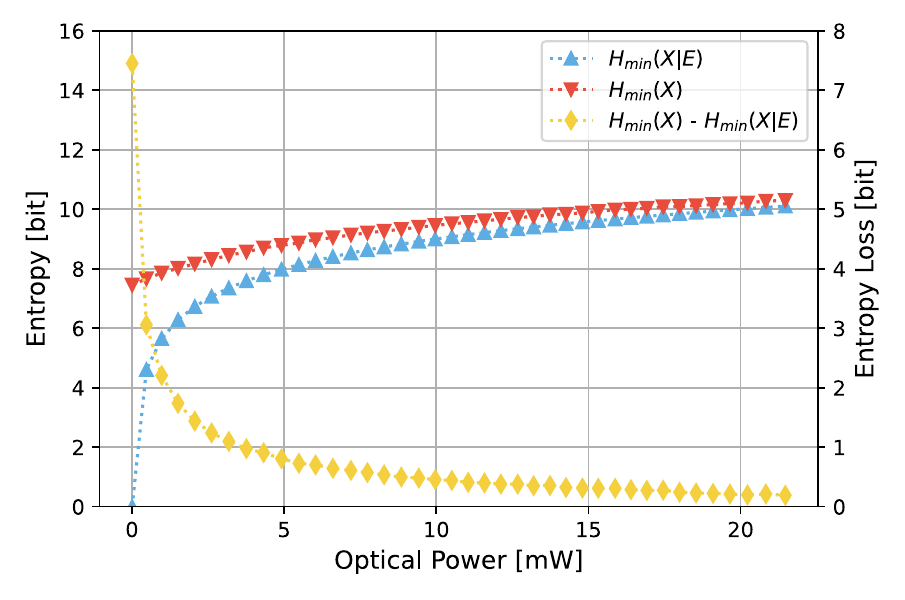}
    \caption{Relation between the LO optical power, the quantum conditional min-entropy $H_{\text{min}}(X|E)$ and the classical min-entropy $H_{\text{min}}\left( X \right)$. Since $H_{\text{min}}\left( X \right)$ upper-bounds the $H_{\text{min}}(X|E)$, also their difference is reported, showing that for increasing LO powers this quantity (entropy loss) reduces to a fraction of a bit.}
    \label{fig:HminQtmCls}
\end{figure}

After the QRNG phase-space resolution $\delta_{q,p}$ has been estimated, the conditional quantum min-entropy was determined for each local oscillator's power $P_{LO}$ used; the results are reported in Fig. (\ref{fig:HminQtmCls}).
In the case of maximum $P_{LO}$ ($\sim 21~\si{\milli\watt}$), $H_{\text{min}}\left( X\vert E \right)$ lower-bounds to $10.106 \pm 0.005$ bit the randomness that can be extracted from a $16$-bit heterodyne measurement ($8$-bit per channel).
Therefore, the achievable secure generation rate is:
\begin{equation}
  R_{sc}=R_{rw}\, H_\text{min} \left(X\vert E\right) = 20.212 \:\: \text{Gbps}\, ,
  \label{eq:secure_gen_rate}
\end{equation}
with $R_{rw}$ the equivalent ADC sampling rate after the {off-line} pre-processing procedure {digital re-sampling} ($2~\si{GSps}$).
Notice that even with low $P_{LO}$, $475 \pm 10\, \si{\micro\watt}$, the device can deliver more than $4~\si{Gbps}$ of secure random bits.
Such sensitivity can be further enhanced with integrated photodetectors \cite{9380443} to eliminate the coupling losses due to hybrid integration.
Additionally, incorporating integrated optical sources or gain chips butt-coupled to the PIC can reduce the coupling losses associated with grating couplers \cite{9380443}.
These modifications could increase the coupling efficiency by at least fourfold, which would correspondingly improve the sensitivity by a comparable amount.

\begin{figure}[t]
    \centering
    \includegraphics[width=\columnwidth]{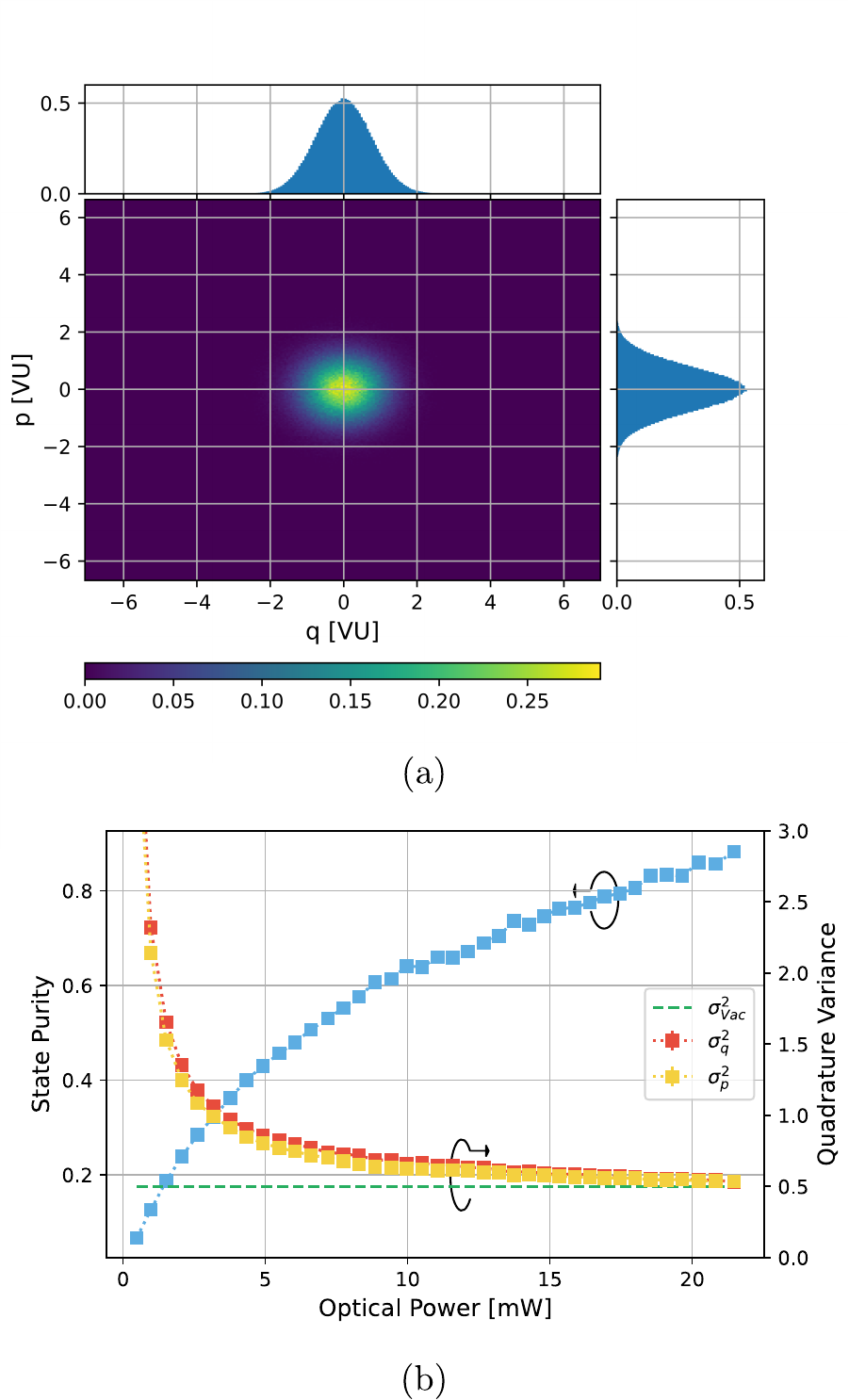}
    \caption{a) Experimental source (vacuum) state tomography with maximum LO power.
    On the axes sides, the associated marginal distributions are reported. b) The graph presents the relation between the LO optical power, the estimated quadratures variances and the purity of the measured state.
    Since the classical noise is considered un-trusted, as the LO power increases, the variances approach the expected vacuum values ($1\text{/}2$) and the purity tends to one.}
    \label{fig:purity}
\end{figure}

For comparison in Fig. (\ref{fig:HminQtmCls}) the un-conditional classical min-entropy $H_{\text{min}}\left( X \right)$, a quantity that upper-bounds $H_{\text{min}}\left( X\vert E \right)$, is also reported.
The gap between the two possesses a decreasing trend with respect to $P_{LO}$, reaching $0.191 \pm 0.005$ bit when the latter is set to the maximum tested power.
Since the vacuum state was injected while running the experiments, the latter discrepancy has a straight forward interpretation. 
Indeed, when calibrating the receiver, its classical noise was deemed as un-trusted.
Consequently, there is an additional contribution to the vacuum's quadrature variances of $1\text{/}2$.
This is exemplified in Fig. (\ref{fig:purity}), where we depicted the state's estimated purity and the corresponding variances of the quadratures $\sigma_{q,p}^2$ relative to the local oscillator power.
As it can be observed, higher levels of $P_{LO}$ corresponds an increase in the purity.
At the highest LO power, the purity peaks at a value of {$0.883 \pm 0.003$} with {$\sigma^2_q = 0.5317 \pm 0.0013$} and {$\sigma^2_p = 0.5347\pm 0.0013$}.
Such a trend stems from the fact that for increasing LO power, the ratio between the quantum signal and the classical noise (clearance) improves.

Following the calibration of the QRNG receiver, approximately $50~\si{GBytes}$ of data was collected for random number extraction.
For this purpose, the $P_{LO}$ was settled to its peak value of $21.15 \pm 0.01~\si{\milli\watt}$ reached during calibration.
The 2-universal hashing function used was the Toeplitz bit-wise matrix-vector multiplication with $n=17600$ columns and $m=11008$ rows.
These numbers guarantee a security parameter $\varepsilon^\prime \sim 10^{-10}$ of the whole hashed data and approximate the ratio $m \text{/} n\sim H_\text{min}\left( X\vert E \right)/2\, n_\text{bit}$ to preserve generation efficiency:
\begin{equation}
 R_{sc}=\left(\dfrac{m}{n}\right)\, 2\, n_\text{bit}\, R_{rw} \approx\, 20.015\:\: \text{Gbps}\, .
\end{equation}
Finally, with our matrix dimension choice, we obtained $2.5~\si{GBytes}$ of hashed data from the original $50~\si{GBytes}$ of raw measurements.
The latter were then analyzed with the standard NIST random test batteries suite; the results are reported in Table (\ref{tab:NIST}).

\begin{table}[]
	\begin{center}
		\begin{tabular}{lcc}
		    \hline\rule{0pt}{3ex}
			Test & p-value & Test Result \\
			\hline
			Frequency & 0.768 & PASSED \\
			BlockFrequency & 0.559 & PASSED \\
			CumulativeSums & 0.386 & PASSED \\
			Runs & 0.533 & PASSED \\
			LongestRun & 0.323 & PASSED \\
			Rank & 0.091 & PASSED \\
			DFT & 0.537 & PASSED \\
			NonOverlappingTemplate & 0.646 & PASSED \\
			OverlappingTemplate & 0.963 & PASSED \\
			Universal & 0.342 & PASSED \\
			ApproximateEntropy & 0.800 & PASSED \\
			LinearComplexity & 0.748 & PASSED \\
			RandomExcursions & 0.234 & PASSED \\
			RandomExcursionsVariant & 0.628 & PASSED \\
			Serial & 0.539 & PASSED \\
			LinearComplexity & 0.809 & PASSED \\
			\hline
		\end{tabular}
	\end{center}
    \captionsetup{font=footnotesize}
	\caption{NIST test results with an hashed file of approximately $2.5~\si{Gbyte}$. The extractor, designed to ensure a security parameter {$\varepsilon^\prime \sim 10^{-10}$} of the whole hased data, is implemented as a Toeplitz matrix-vector modulo two multiplication with column and row dimensions respectively of $17600$ and $11008$.}
    \label{tab:NIST}
\end{table}

\section{Conclusions}
This work presents, to the best of our knowledge, the first source-device independent QRNG scheme on a chip and currently the fastest among the Semi-DI class, capable of a secure generation rate greater than $20~\si{Gbps}$.
Such a result was possible by exploiting integrated photonics alongside fast electronics.
Indeed, the latter combination allowed us to achieve a high generation efficiency, ensuring several Gbps even with the local oscillator power well below $1~\si{\milli\watt}$.
The generator was developed to promote system designs that are simpler, compact, and reliable.
This is achieved through the 4x4 MMI interferometer upon which our heterodyne receiver is based.
Indeed, no active feedback control loops are required to ensure the orthogonality of the measured quadratures.
Moreover, the 4x4 MMI, being a fully passive structure, should make the optical receiver resistant to harsh operating conditions.
Therefore, with the potential qualities mentioned before, the proposed solution should represent an appealing candidate to address the challenges posed by secure space applications.
Especially the ones that also require high-performance operations, such as satellite-based QKD.
Indeed, given the demonstrated secure generation rate obtained in this paper, our module could power satellite QKD transmitters with a repetition rate higher than $5~\si{\giga\hertz}$.

To conclude, our QRNG combines high security and record-high performance with the benefits of an integrated platform.
We believe that this is a significant step forward for the exploitation of this technology in demanding real-world scenarios such as high-speed QKD, either ground or space based.

\begin{acknowledgments}
    We acknowledge the Italian Space Agency for support.
    This work was funded by the project QRNG of the Italian Space Agency (ASI, Accordo n. 2017-31-H.0): \virgolette{Realizzazione integrata di un generatore quantistico di numeri casuali}. We thank the CloudVeneto facility for compuRtational resources. We also thank CAPRI (University of Padova Strategic Research Infrastructure Grant 2017: “CAPRI: Calcolo ad Alte Prestazioni per la Ricerca e l’Innovazione”) for computational resources.  
\end{acknowledgments}

\end{document}